\documentclass[preprint,showpacs,preprintnumbers]{revtex4}
\usepackage{amssymb}

\usepackage{graphicx}
\usepackage{dcolumn}
\usepackage{bm}

\input{tcilatex}

\begin{document}

\title{A spatiotemporal-chaos-based cryptosystem taking advantages of both
synchronous and self-synchronizing schemes}
\author{Huaping L\"{u}$^{1,2},$ Shihong Wang$^{1},$ Xiaowen Li$^{1},$
Guoning Tang$^{1},$ Jinyu Kuang$^{3},$ Weiping Ye$^{3}$, Gang Hu$^{4,1,5\ast
}$}
\affiliation{$^{1}$Department of Physics, Beijing Normal University, Beijing 100875, China%
\\
$^{2}$Department of Physics,Xuzhou Normal University, Xuzhou 221009,China}
\affiliation{$^{3}$Department of Electronics, Beijing Normal University, Beijing
100875,China\\
$^{4}$China Center for Advanced Science and Technology (CCAST) (World
Laboratory), P.O.Box 8730, Beijing 100080, China}
\affiliation{$^{5}$The Key Laboratory of Beam Technology and Material Modification of
Ministry of Education,Beijing Normal University, Beijing 100875, China\\
$^{\ast }$Correspondent author (Email: ganghu@bnu.edu.cn)}
\date{\today }

\begin{abstract}
Two-dimensional one-way coupled map lattices are used for cryptograph where
multiple space units produce chaotic outputs in parallel. One of the outputs
plays the role of driving for synchronization of the decryption system while
the others perform the function of information encoding. With this
separation of functions the receiver can establish a self-checking and
self-correction mechanism, and enjoys the advantages of both synchronous and
self-synchronizing schemes. A comparison between the present system with the
system of Advanced Encryption Standard, AES, is presented in the aspect of
channel noise influence. Numerical investigations show that our system is
much stronger than AES against channel noise perturbations, and thus can be
better used for secure communications with large noise channel exists in
open channels, \textit{e.g}., mobile-phone secure communications.
\end{abstract}

\maketitle

Since the pioneering work by Pecora and Carrol, chaos-based secure
communication has attracted great attention \cite%
{Hu1,Hu2,Hu3,Hu4,Hu5,Hu6,Hu7,Hu8,Hu9,Hu10}. In the first stage, chaos
communication focused on how to hide secret message in chaotic signals and
how to establish communication links between transmitters and receivers
through various methods of chaos synchronization. After its development of a
decade, now any new valuable designs of chaotic cryptosystems should be
examined from the standard evaluations of overall cryptographic properties
popularly used in conventional secure communications. Among these
properties, the following three criterions are the most important: security;
efficiency of performance; and robustness against channel noise
(realizability) \cite{Hu11,Hu12,Hu13}. In this paper we will focus the
property of resistance against noise disturbances in transmission channels,
this property is closely related to the cryptographic methods and schemes.

\ For performing secure communications various encryption methods, including
conventional methods and chaos-based ones, have been proposed of which all
stream cipher cryptosystems can be classified to two schemes, synchronous
scheme and self-synchronizing scheme. Each scheme has its own advantages as
well as its disadvantages. Let us consider chaos-based cryptosystems for
explaining this matter.

With synchronous scheme both transmitter and receiver use identical
autonomous chaotic systems for encryption and decryption transformations.
The synchronization between the transmitter and the receiver is achieved by
choosing same initial conditions, and the ciphertexts transmitted in the
open channel plays no role for chaos synchronization. In this case, the
initial values of chaotic variables often serve as secret the key for the
security. A serious problem of this scheme is that the synchronization
between the transmitter and the receiver, $i.e.$, the realizability of the
communication, is unstable against certain mismatches between the both sides
in communication. In particular, this scheme has the following
disadvantages. (i) To be resistant against plaintext-known attacks, one has
to use different initial conditions for different times of communications, $%
i.e.$, use $\mathbf{x}_{0}=\mathbf{a}+\mathbf{b,x}_{0}=(x_{1},x_{2},\cdot
\cdot \cdot ,x_{m}),$ as the initial values of the dynamical variables where 
$\mathbf{a}=(a_{1,}a_{2},\cdot \cdot \cdot ,a_{m})$ is kept unknown for
public as private the key which is unchanged for different communications
while $\mathbf{b=}(b_{1},b_{2},\cdot \cdot \cdot ,b_{m})$ is changed in
different communications and it is transmitted in open channel. During the
open channel transmission any bit error of $\mathbf{b}$ can cause the
collapse of chaos synchronization between the transmitter and the receiver
and definitely destroy the whole communication. (ii) With synchronous
scheme, both transmitter and receiver perform parallel chaotic motions, any
encryption and decryption transformations should be adjusted in synchronized
time. Therefore, any missing of bits or packets of bits in transmitted
ciphertexts may break the synchronization. (iii) Any temporary perturbations
in encryption and decryption systems can definitely destroy the whole
communication, and this breakdown cannot be overcome during the same
communication. Nevertheless, synchronous scheme has some advantages
balancing its disadvantages. A remarkable advantage is that the synchronous
scheme has no bit error avalanche effect. The synchronization between the
transmitter and the receiver is realized by taking identical initial
conditions, and the transmitted ciphertexts do not play the role of driving
for synchronization of the receiver. Thus, one bit error in the ciphertexts
causes one bit error only in the received plaintext, and the system can be
rather robust against strong noise perturbations in the communication
channel.

With self-synchronizing scheme, the encryption dynamics of the transmitter
is chaotic, and the signal transmitted in the open channel now plays double
roles of information carryer (ciphertexts) and the receiver driving for
chaos synchronization. The receiver does not take (and actually does not
know) the initial condition of the transmitter, and it reaches
synchronization to the transmitter from arbitrary initial condition through
the driving of the transmitted signal. In this case, the self-synchronizing
scheme enjoys desirable advantages of stability of communication. It does
not need to transmit initial condition; and it can automatically recover
chaos synchronization and successful communication after any breaks due to
bit or bit packet missing, or due to temporary disorders of the transmitter
and the receiver, because the subsequent driving signal can resynchronize
the receiver and avoid collapse of communication. However, the
self-synchronizing scheme has its own disadvantages as well. An error in the
driving bit may cause a large number of error bits in the received
plaintexts, this is the so-called bit error avalanche effect. With this
effect, the self-synchronizing scheme becomes weak in resistance against
channel noise, and the robustness of self-synchronizing secure
communications in strong noise environment is usually much worse than the
secure communications of synchronous scheme.

All cryptosystems (both chaos-based and conventional ones) can be classified
to one of the above two schemes. It is noticed that typical block cipher
systems like Advanced Encryption Standard, AES, can be regarded to belong to
the self-synchronizing scheme because in block cipher systems the
ciphertexts are used as the driving signal of the inverse computations of
the receiver, and they enjoy the advantages of stability of the
self-synchronizing scheme as well as suffer from the bit error avalanche
disadvantage of the same scheme.

The central task of the present paper is to suggest, for the first time to
our knowledge, a cryptosystem based on spatiotemporal chaos which combines
the advantages of both synchronous and self-synchronizing schemes and avoid
their disadvantages. In \cite{14} we proposed to use spatiotemporally
chaotic cryptosystems to reach high security; in \cite{15} we suggested to
simultaneously use many chaotic outputs from different space units to
provide high encryption (decryption) efficiency; and in \cite{16} we \
suggested to combine the methods of \cite{14} and \cite{15} to achieve
optimal overall cryptographic properties, $i.e.$, high security and fast
encryption (decryption) speed. In all the above works we used clearly
self-synchronizing cryptographic scheme. In the present paper we will go
further from \cite{14} and \cite{16} by separating the transmitted signals
to two distinctive parts, one driving signal for the function of
synchronizing the receiver and the other ciphertext signal for the function
of information carryer (note, in the typical self-synchronizing scheme both
functions are performed simultaneously by the same transmitted signal). With
this separation of functions, our cryptosystem enjoys all the advantages of
both synchronous and self-synchronizing schemes and avoids their own
disadvantages.

We use the following one-way coupled map lattice \cite{16} for

Encryption transformation:

\ \ \ \ \ \ \ \ \ \ \ \ \ \ \ \ \ \ 
\begin{eqnarray}
x_{n+1}(j) &=&(1-\varepsilon )f_{j}[x_{n}(j)]+\varepsilon f_{j-1}[x_{n}(j-1)]
\nonumber \\
f_{0}(x) &=&4x(1-x)  \TCItag{1a} \\
f_{j}(x) &=&a_{j}x(1-x),\text{ \ \ }j=1,2,...,m  \nonumber
\end{eqnarray}

\begin{center}
\begin{eqnarray}
z_{n+1}(0,0) &=&(1-\varepsilon )f[z_n(0,0)]+\varepsilon f_m[x_n(m)] 
\nonumber \\
y_n(0,0) &=&z_n(0,0)\times 2^h\ \ \ \func{mod} 1  \TCItag{1b} \\
f(x) &=&4x(1-x)  \nonumber
\end{eqnarray}

\begin{eqnarray}
y_{n+1}(1,0) &=&(1-\varepsilon )f[y_{n}(1,0)]+\varepsilon f[y_{n}(0,0)] 
\nonumber \\
y_{n+1}(0,1) &=&(1-\varepsilon )f[y_{n}(0,1)]+\varepsilon f[y_{n}(0,0)] 
\nonumber \\
y_{n+1}(j_{1},0) &=&(1-\varepsilon )f[y_{n}(j_{1},0)]+\varepsilon
\{0.8f[y_{n}(j_{1}-1,0)+0.2f[y_{n}(j_{1}-2,0)]\},\ \   \nonumber \\
\ j_{1} &=&2,...,N  \nonumber \\
y_{n+1}(0,j_{2}) &=&(1-\varepsilon )f[y_{n}(0,j_{2})]+\varepsilon
\{0.2f[y_{n}(0,j_{2}-1)+0.8f[y_{n}(0,j_{2}-2)]\},\ \   \nonumber \\
\ j_{2} &=&2,...,N  \TCItag{1c} \\
y_{n+1}(j_{1},j_{2}) &=&(1-\varepsilon )f[y_{n}(j_{1},j_{2})]+\varepsilon
\{0.5f[y_{n}(j_{1}-1,j_{2})\ +0.5f[y_{n}(j_{1},j_{2}-1)]\},\   \nonumber \\
2 &\leqslant &j_{1}+j_{2}\leqslant N  \TCItag{1d}
\end{eqnarray}
\end{center}

\begin{eqnarray}
S_{n}(j_{1},j_{2}) &=&[K_{n}(j_{1},j_{2})+I_{n}(j_{1},j_{2})]\ \ \text{mod }%
2^{\nu },\ \   \nonumber \\
\ K_{n}(j_{1},j_{2}) &=&[\text{int}(y_{n}(j_{1},j_{2})\times 2^{\mu }]\ \ 
\text{mod }2^{\nu },\ \ 2\leqslant j_{1}+j_{2}\leqslant N  \TCItag{2a}
\end{eqnarray}

\begin{equation}
D_{n}=[\text{int}(y_{n}(N/2,N/2)\times 2^{\mu }]\text{ \ \ mod 2}^{\nu } 
\tag{2b}
\end{equation}

\[
x_{n}(0)=D_{n}\times 2^{-\nu } 
\]%
where parameter $a_{j}$, $j=1,2,\cdot \cdot \cdot ,m$ are kept secret,
serving as the private key, $S_{n}(j_{1},j_{2})$ are ciphertexts, carrying
information of plaintexts, and $D_{n}$ serves as the driving signal. Both $%
S_{n}(j_{1},j_{2})$ and $D_{n}$ are transmitted in the public channel to the
receiver. For specific we will, throughout the paper, take parameters $h=26$,%
$\ \mu =58$,$\ \nu =32$, $\ \varepsilon =0.99$, $\ m=3$, $\ N=6$, and $%
a_{1}=a_{2}=a_{3}=3.7$.

The decryption transformation of the receiver can be obtained symmetrically
by replacing $x_{n}(j)$, $z_{n}(0,0)$, $y_{n}(j_{1},j_{2})$, the key $%
\mathbf{a=(}a_{1},a_{2},...,a_{m})$, with $x_{n}^{\prime }(j)$, $%
z_{n}^{\prime }(0,0)$, $y_{n}^{\prime }(j_{1},j_{2})$, and the test key $%
\mathbf{b=(}b_{1},b_{2},...,b_{m})$, respectively, and with

\begin{equation}
x_{n}^{^{\prime }}(0)=x_{n}(0)=D_{n}/2^{\nu }  \tag{3}
\end{equation}%
\ or\ \ \ \ \ \ \ \ \ \ 
\[
x_{n}^{^{\prime }}(0)=D_{n}^{^{\prime }}/2^{\nu } 
\]%
Moreover, the plaintext can be extracted by the receiver as

\begin{eqnarray}
I_{n}^{\prime }(j_{1},j_{2}) &=&[S_{n}(j_{1},j_{2})-K_{n}^{\prime
}(j_{1},j_{2})]\ \ \text{mod }2^{\nu },\ \ \   \nonumber \\
\text{ \ \ }K_{n}^{\prime }(j_{1},j_{2}) &=&[\text{int}(y_{n}^{\prime
}(j_{1},j_{2})\times 2^{\mu }]\ \ \text{mod }2^{\nu }  \TCItag{4a}
\end{eqnarray}

\begin{equation}
D_{n}^{^{\prime }}=[\text{int}(y_{n}^{\prime }(N/2,N/2)\times 2^{\mu }]\text{
\ \ \ mod 2}^{\nu }  \tag{4b}
\end{equation}%
By taking the secret key identical to the transmitter $%
b_{j}=a_{j},j=1,2,...,m,$ the receiver can reach synchronization with the
transmitter, and then correctly receive the private message as

\begin{eqnarray}
y_{n}^{^{\prime }}(j_{1},j_{2}) &=&y_{n}(j_{1},j_{2}),\text{ \ }%
K_{n}^{\prime }(j_{1},j_{2})=K_{n}(j_{1},j_{2})\text{,\ \ }D_{n}^{^{\prime
}}=D_{n}\ \   \nonumber \\
\text{ }I_{n}^{\prime }(j_{1},j_{2}) &=&I_{n}(j_{1},j_{2}),\ \ 2\leqslant
j_{1}+j_{2}\leqslant N  \TCItag{5}
\end{eqnarray}

\bigskip

The schematic encrypting structure is given in Fig.1(a). First, we use a 1D
lattice chain of length $m=3$ (three circles) for setting the secret keys $%
\mathbf{a=(}a_{1},a_{2},a_{3})$ in Eq.(1a). This arrangement can guarantee
full sensitivity of the encryption processes to all key parameters on one
hand, and reduce as much as possible the number of noncipher sites on the
other hand. Second, following the 1D chain is a 2D one-way coupled map
lattice of Eqs.(1c) and (1d), producing $W=25$ chaotic sequences (the square
sites of Fig.1(a)), in parallel for fast encryption. The 2D structure is for
reducing the system length and thus effectively reducing synchronization
time and the associated error bit avalanches. And the nearest neighbor and
next to the nearest neighbor coupling structure of Eq.(1c) is for enormously
increasing the cost of any inverse analytical computation attacks. Third,
there are two mod operations Eqs.(1b) and (2a) which may considerably
enhance the sensitivity of chaos synchronization to the key parameter
variations, and thus reach high practical encryption security. Finally, all
the square maps output signals in parallel of which the $W=24$ empty squares
encrypt plaintexts $\left[ S_{n}(j_{1},j_{2})=K_{n}(j_{1},j_{2})\oplus
I_{n}(j_{1},j_{2}),\text{ see Eq. (2a)}\right] $ and \ produce ciphers, and
the single black square produces driving signal $D_{n}$\ which does not
encrypt any plaintext and feeds back to drive the coupled maps lattice by $\
x_{n}(0)=D_{n}/2^{\nu }.$ The decryption transformation of the receiver has
similar structure to Fig.1(b), where the driving and decrypting are
organized by $\ x_{n}^{^{\prime }}(0)=D_{n}/2^{\nu },$\ $I_{n}^{\prime
}(j_{1},j_{2})=S_{n}(j_{1},j_{2})\ominus K_{n}^{\prime }(j_{1},j_{2})$\ as
defined in Eq.(4a). With $b_{j}=a_{j},j=1,2,3$ we can obtain spatiotemporal
chaos synchronization between the tranmitter and the receiver, and thus the
receiver can obtain $I_{n}^{\prime }(j_{1},j_{2})=I_{n}(j_{1},j_{2}).$

There are two crucial key points of our cryptosystem should be emphasized.
First, in the transmitter we use signal $D_{n}$ for the feedback $\
x_{n}(0)=D_{n}/2^{\nu }.$ Since $D_{n}$ does not make encrypting, the
encryption system is autonomous spatiotemporal chaos (with conventional
self-organizing scheme, the encryption system is nonautonomous chaos with
plaintext involving in the dynamics). Second, in the receiver the two
functions of ciphertexts and synchronization driving, which are usually
played by the same transmitted signal simultaneously in self-synchronizing
scheme, are now played separately by two different parts of transmitted
bits: $S_{n}(j_{1},j_{2})$ from the empty squares serve as the ciphertexts
(which play no role of driving) while $D_{n}$ from the black square serves
as the driving signal synchronizing the decryption system (which does not
produce ciphers). The most significant point is that the decryption system
can produce a driving signal $D_{n}^{^{\prime }}$ by its own dynamics [see
Fig.1(b)], which must be identical to the transmitted driving signal $D_{n}$%
, $D_{n}^{^{\prime }}=D_{n,}$ if chaos synchronization between the
transmitter and the receiver is achieved. This identity can be used by the
receiver to check the validity of the communication and to correct errors
whenever the transmitted driving $D_{n}$ is disturbed by noise in the open
channel, as we will see later. This self-checking and self-correcting
capacity of the receiver is a great advantage of our cryptosystem over all
known methods of both synchronous and self-synchronizing schemes.

With the above mechanism our cryptosystem enjoys all the advantages of both
self-synchronizing and synchronous schemes, and avoid their disadvantages.
On one hand, we do not need to transmit initial condition for each
communication, and do not suffer from the collapse caused by bit and bit
packet loss or by temporary disturbances of the transmitter and the receiver
machines. On the other hand, our system enjoys all the advantages of
synchronous scheme as well. In particular, the bit errors of the transmitted
driving signal can be checked and corrected, and the bit error avalanche
caused by the open channel noise can be practically eliminated. The present
cryptosystem has strong robustness against channel noise. These conclusions
will be fully confirmed by the following numerical computations.

First let us specify the detail of numerical simulations of our system. We
use two computers, one serves as the transmitter and the other as the
receiver, for simulating chaos-based secure communications by using
Eqs.(1)--(5) with the given parameters for simulations. The plaintexts $%
I_{n}(j_{1},j_{2})$ are arbitrarily chosen. In each communication the
initial condition of the transmitter is chosen as random numbers, and
different communications take different initial conditions. The receiver
takes initial conditions also randomly, which are different from those of
the transmitter. Starting each communication, the transmitter first send
driving bits $D_{n}$ to the receiver only for about 30 iterations in our
case (slight less than 1K bits), these driving signal bits are enough to
synchronize the decryption system. After this transient, the transmitter
send both ciphertexts $S_{n}(j_{1},j_{2})$ and driving $D_{n}$ in the same
open channel, and then the standard secure chaotic communication works. We
simulate noise disturbance in the open channel by certain bit error rate of
transmitted signal that each transmitted bit has a probability $P$ to
alternate its value from 1 to 0 (also from 0 to 1). During the normal
communication the receiver can continually receive the transmitted driving
signal $D_{n}$ and autonomously produce its own driving signal $%
D_{n}^{\prime }$. The receiver can compare $D_{n}^{\prime }$ with $D_{n},$
and monitor the bit errors between $D_{n}$ and $D_{n}^{^{\prime }}.$ Suppose
the number of error bits between $D_{n}$ and $D_{n}^{\prime }$ is $E_{n},$
the receiver can distinguish three different situations. (i) $E_{n}=0$, the
normal communication will be continued since full synchronization is kept.
(ii)

\begin{equation}
0<E_{n}<A,\text{ \ \ \ \ }A\ll \nu  \tag{6}
\end{equation}%
then most of bits of $D_{n}$ and $D_{n}^{^{\prime }}$ coincide with each
other, indicating synchronization between the transmitter and the receiver,
and a small number of nonidentical bits are due to the $D_{n}$ bit errors
caused by the channel noise. In this case, the receiver can use $%
D_{n}^{^{\prime }}$ replacing $D_{n}$ to drive the decryption system for
avoiding bit error avalanche of $D_{n}$. Here $A$ is chosen to be
''sufficiently small'' that sufficiently many identical bits of $D_{n}$ and $%
D_{n}^{^{\prime }}$ convince the chaos synchronization of the transmitter
and the receiver, and ''sufficiently large'' that the affect of channel
noise can be effectively eliminated. In our simulations we take $A=8$ for
the total $\nu =32$ bits of $D_{n}$ (also $D_{n}^{^{\prime }})$. (iii) If $%
E_{n}>A,$ it is assumed that the synchronization between the transmitter and
the receiver may be broken due to, $e.g.$, bit packet missing or some
temporary disturbance of the transmitter or receiver machines, we thus take $%
D_{n}$ as the driver of the decryption system for recovering the chaos
synchronization.

In summary, with condition (ii) [Eq.(6)] we judge successful chaos
synchronization between the transmitter and the receiver with the
transmitted driving signal having certain error bits caused by channel
noise, and then we use $D_{n}^{^{\prime }}$ as driving for avoiding bit
error avalanche; with condition (iii) it becomes possible that chaos
synchronization may be broken due to some reasons, we thus use $D_{n}$ as
driving for synchronization and avoiding possible communication collapse;
and with condition (i) chaos synchronization and normal communication are
guaranteed, we can use either $D_{n}$ or $D_{n}^{^{\prime }}=D_{n}$ for
driving.

Of course, chaos synchronization can be broken even in the case of (ii). The
largest probability for this event to happen once can be easily computed by
the error function

\begin{eqnarray}
Q_{1} &\approx &\frac{1}{\sqrt{2\pi \nu \sigma }}\int_{\nu -A}^{\nu }e^{-%
\frac{(x-\frac{\nu }{2})^{2}}{2\nu \sigma }}dx\approx \frac{1}{\sqrt{2\pi
\nu \sigma }}e^{-\frac{(\frac{\nu }{2}-A)^{2}}{2\nu \sigma }}  \nonumber \\
\sigma &=&\frac{1}{4}  \TCItag{7}
\end{eqnarray}%
with

\begin{equation}
\nu =32,\text{ \ \ }A=8  \tag{8}
\end{equation}%
we have $Q_{1}\approx 2\times 10^{-3}.$ And the largest probability for
these events to happen successively $J$ times is

\begin{equation}
Q_{J}=Q_{1}^{J}\approx 2^{J}\times 10^{-3J}  \tag{9}
\end{equation}%
which goes to zero very quickly as J increases (actually, J=2 is already
hardly observable). Therefore, chaos synchronization breaking due to any
reasons can be found almost immediately, and synchronization can be
reestablished easily with the $D_{n}$ driving.

On the other hand, with function (ii) our system is very strong in
resistance against channel noise. Suppose the bit error probability of $%
D_{n} $ caused by channel noise is $P.$ With Eq.(6) this noise has
probability $q$ to cause avalanche

\begin{eqnarray}
q &=&\sum_{k=A+1}^{\nu }C_{\nu }^{k}P^{k}(1-P)^{\nu -k}\approx
\sum_{k=A+1}^{\nu }C_{\nu }^{k}P^{k}e^{-P(\nu -k)}  \nonumber \\
&=&e^{-P\nu }\sum_{k=A+1}^{\nu }C_{\nu }^{k}P^{k}e^{Pk}  \TCItag{10}
\end{eqnarray}%
Let us define $R$ as the error rate of the received plaintext

\begin{equation}
R=\frac{M_{e}}{M}  \tag{11}
\end{equation}%
with $M$ being the total number of transmitted ciphertext bits and $M_{e}$
being the error bits among these $M$ bits. If we have

\begin{equation}
R\approx P  \tag{12}
\end{equation}%
the avalanche effect can be regarded to be ineffective and our chaos-based
secure communication is as strong as normal nonsecure communication in
resistance against channel noise. With (8) and (10), Eq.(12) is valid up to $%
P\approx 1/30$ (then we have $q<\sum_{n=9}^{32}\frac{1}{n!}\approx 2.5\times
10^{-6}$ probability for a bit to cause an avalanche and a single avalanche
can cause about $9\times 10^{3}$ error bits in our system).

For comparison we also simulate secure communication by applying AES
cryptosystem. This system has an avalanche of 128 bits, $i.e.$, any bit
error in the transmitted ciphertext can cause 128-bit errors in the received
plaintext. As a complete comparison we should investigate all the properties
of security, performance and robustness of the cryptosystems. The security
problem has been studies in detail in Refs.\cite{14} and \cite{16} ,
performance problem is analyzed in great detail in \cite{16}for software
realization. Our chaos-based cryptosystem has encryption (decryption) speed
2 to 3 times faster than that of AES with 128-bit key \cite{17} because the
former takes parallel outputs and one-round encryption structure. We have
confirmed this conclusion by using different computers and by taking windows
2000 operating system and C compiler\ (in comparisons, both cryptosystems
were computed with same computers). In the present work we focus on the
comparison of robustness of both systems against channel noise, simulated by
bit error probability of the transmitted signal.

In Fig.2 we plot the error rate of the received plaintext $R$ against the
channel noise of error probability $P$ for our spatiotemporal-chaos-based
cryptosystem. For each plot we take $M=10^{5}$ and $<R>$ is an average of
1000 simulating communications. Complete information loss is identified when 
$<R>=0.5$. It is remarkable that difference between our secure communication
with regular nonsecure communication ($<R>$ $\gg P)$ can be observed only
after $P>\frac{1}{30}$ at which any normal (nonsecure) voice communication
is no longer valid. Thus, in our case the operations of information
encryption and decryption do not practically reduce the robustness of
communications.

In Fig.3 we do the same as Fig.2 with the AES system investigated. For small 
$P$, $<R>$ vs. $P$ curves show linear relation with slopes approximately
equal to 64. This behavior is well expected based on its 128-bit error
avalanche. The curve saturates to $<R>\thickapprox 0.5$ at $p\thickapprox
0.03,$ this is sharply different from Fig.2 where the saturation to $%
<R>\thickapprox 0.5$ occurs at about $p\thickapprox 0.1$. And the AES system
cannot be used (unless additional bits are used for error crrections) for
voice communication for $P>10^{-3}$ at which we have $<R>\approx \frac{1}{20}%
,$ breaks of communications but our chaos-based system has $<R>\approx \frac{%
1}{30}$ up to $P\approx \frac{1}{30}.$ It is clear that our
spatiotemporal-chaos-based cryptosystem is much stronger than AES in
resistance against channel noise.

In the investigations of Fig.2 and Fig.3, we consider only bit errors caused
by channel noise, and do not consider influences of bit (or bit-block) loss
and temporary disturbances of the encryption and decryption systems. These
errors may cause some temporary communication breaks for both chaotic and
AES systems. However, the drivings in these systems (driving signal $D_{n}$
for Eqs.(1) and ciphertexts for AES) can well prevent these breaks
developing to communication collapses. And this is a desirable advantage of
both systems over any cryptosystems of synchronous scheme.

If we consider secure communications in network environment, both
synchronous scheme and self-synchronizing scheme can work stably because the
noise strength and the disturbance probability of computers are very weak.
However, communications with city phone lines or with wireless channels must
deal with much stronger noise perturbations and larger error probability in
communicators. In this case, the large error avalanche of the
self-synchronizing scheme and the possible collapse instability of the
synchronous scheme may cause some serious problems. In the present paper, we
have suggested a spatiotemporally chaotic cryptosystem to practically
eliminate the bit error avalanche problem on one hand and to avoid possible
collapse of communication on the other hand, by separating the two functions
of synchronization driving and plaintext encoding to different parts of
transmitted signal. In doing so, all we have\ payed is to reduce the
encryption and transmission efficiency by 4\% ($i.e$., the driving signal $%
D_{n}$ does not produce ciphers which contains 4\% of the total transmitted
bits). And this efficiency loss can be further reduced by decreasing the
number of the driving bits. We expect that our cryptosystem can be used in
secure communication applications where large noise exists in the open
channel, one of such applications may be secure communications with mobile
phones.

This work was supported by the National Natural Science Foundation of China
under Grant No. 10175010, and by Nonlinear Science Project.

\strut

\bigskip

\bigskip

\bigskip

\bigskip

\bigskip

\bigskip

\begin{center}
{\LARGE Figure Captions}
\end{center}

Fig.1 Schematic figure of encryption system (a), and decryption system (b).

Fig.2(a) The average error rate $<R>$ defined in Eq.(11) against the bit
error probability $P$ in the open channel for our chaotic cryptosystem
(1)--(4). For small $P$ we have $<R>\thickapprox P,$ and the bit error
avalanche effect is practically avoided. This avalanche effect can be
clearly observed only after $P\gtrsim \frac{1}{30}$ at which normal
(nonsecure) voice communications may not work. (b) The blowup of a part of
(a). Note: normal scales of $<R>$ are used in (b), unlike (a) where
logarithmic scales are used in both axes.

Figs.3(a),(b) The same as Figs.2(a) and (b), respectively, with AES
investigated. It is clear that the avalanche problem exists for arbitrary $P$%
, and voice communications cannot work for $P>5\times 10^{-4}$ (at which we
have $<R>\approx \frac{1}{30}).$

\ \ 

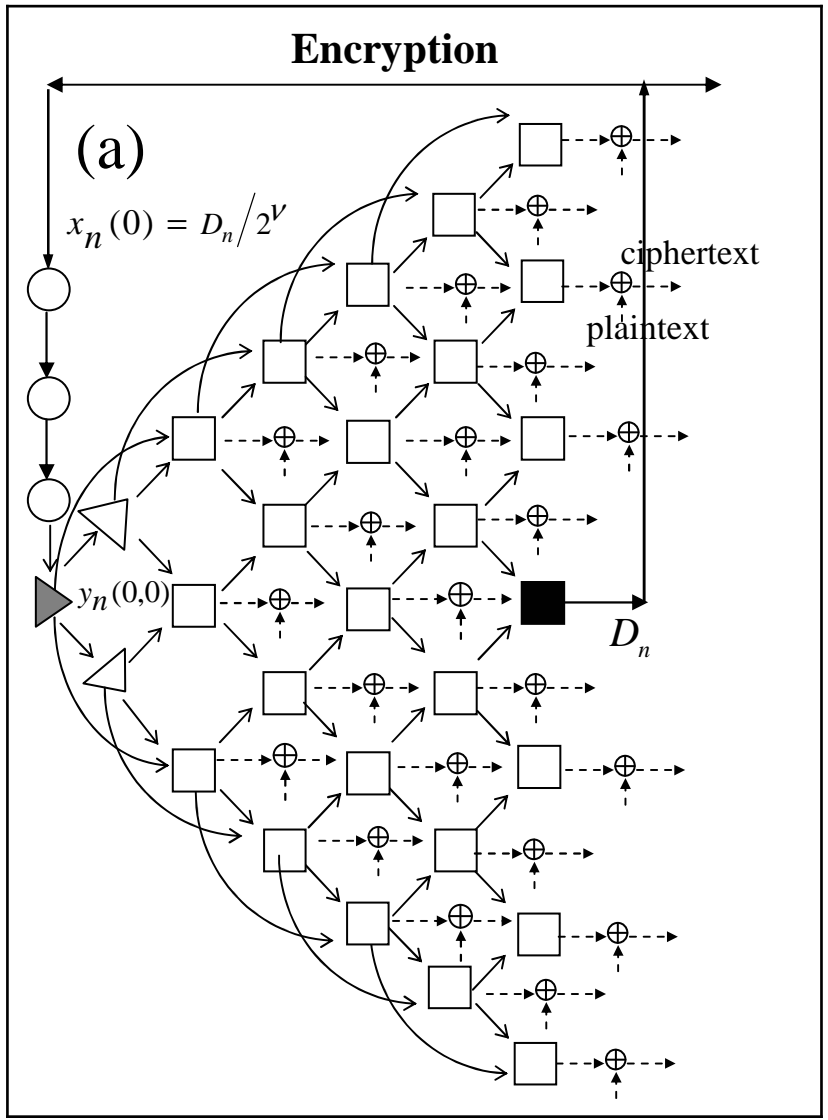
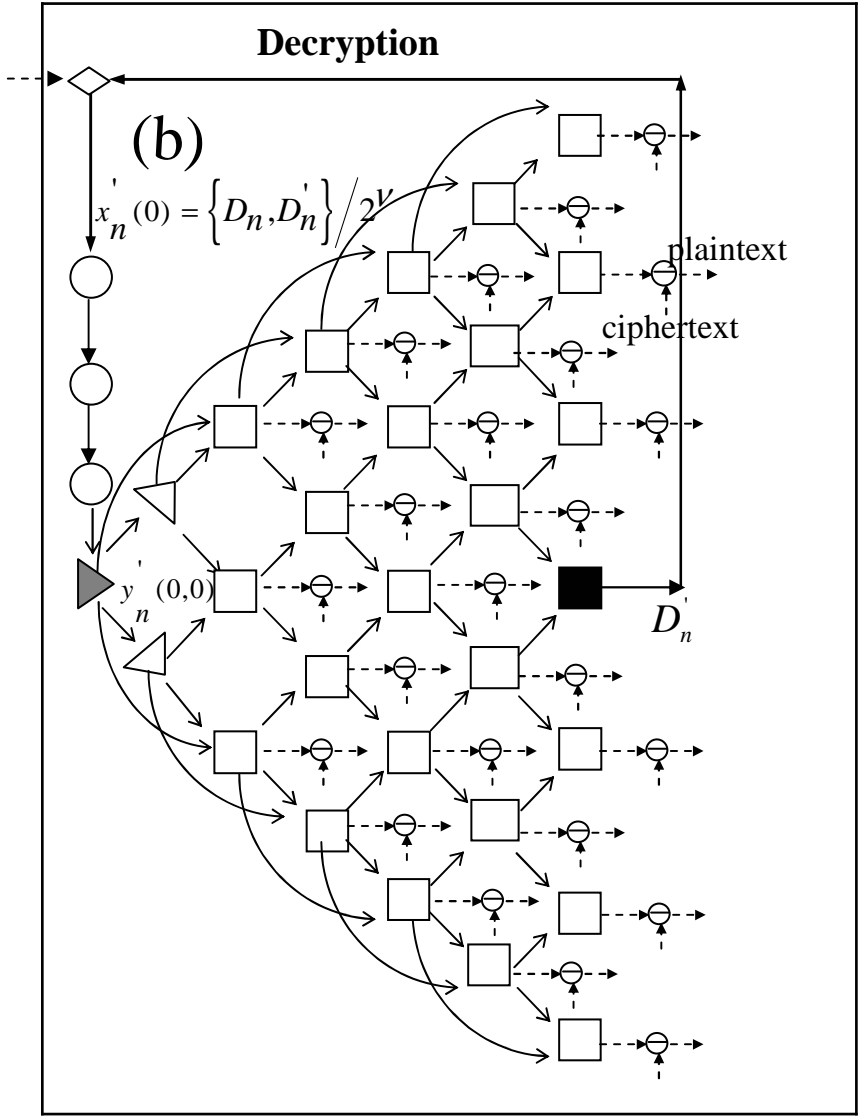
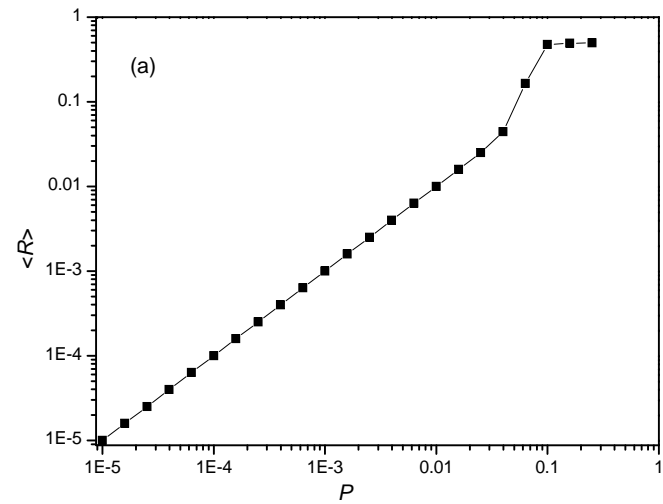
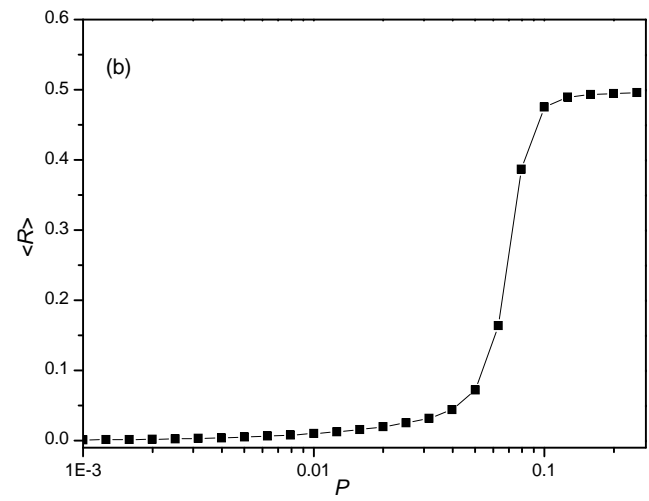
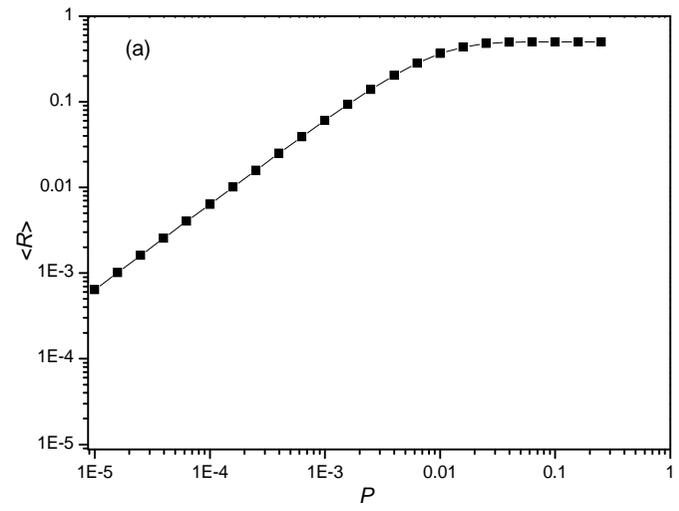
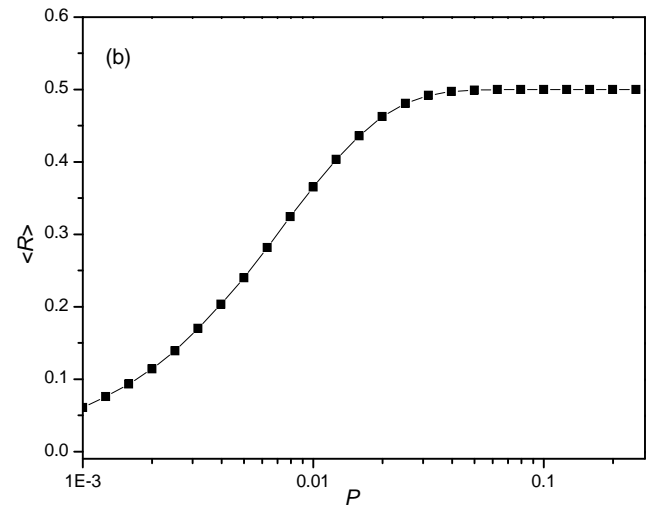
\end{document}